\documentclass[12pt]{article}
\usepackage{amssymb,amsmath,cite,bm}
\usepackage[dvips]{graphicx,color}
\usepackage{psfrag,subfigure}
\newtheorem{theorem}{Theorem}
\newtheorem{remark}{Remark}
\newtheorem{proposition}{Proposition}

\newtheorem{lemma}{Lemma}

\begin{document}

\title{\bf Control and Simulation of Motion of Constrained Multibody Systems Based on Projection Matrix Formulation}

\author{Farhad Aghili\thanks{email: faghili@encs.concordia.ca}}

\date{}

\maketitle

\begin{abstract}
This paper presents a unified approach for inverse and direct
dynamics of constrained multibody systems that can serve as a basis
for  analysis, simulation, and control. The main advantage of the
dynamics formulation is that it does not require the constraint
equations be linearly independent. Thus, a simulation may proceed
even in the presence of redundant constraints or singular
configurations and a controller does not need to change its
structure whenever the mechanical system changes its topology or
number of degrees of freedom. A motion control scheme is proposed
based on a {\em projected inverse-dynamics} scheme which proves to
be stable and minimizes the weighted Euclidean norm of the actuation
force. The projection-based control scheme is further developed for
constrained systems, e.g. parallel manipulators, which have some
joints with no actuators (passive joints). This is complemented by
the development of constraint force control. A condition on the
inertia matrix resulting in a decoupled mechanical system is
analytically derived that simplifies the implementation of the force
control. 
\end{abstract}

\section{Introduction}

Many robotic systems are formulated as multibody
systems with closed-loop topologies, such as manipulators with
end-effector constraints
\cite{Aghili-2005,Khatib-1987,Aghili-Parsa-2009b,Raibert-Craig-1981,Yoshikawa-1987,Aghili-2003,
Yoshikawa-Sugie-Tanaka-1988,Piedboeuf-deCarufel-Aghili-Dupuis-1999,McClamorch-Wang-1988,Aghili-Piedboeuf-2003a,Aghili-Piedboeuf-2006}, cooperative
manipulators \cite{Hu_Golenberg-1993,Namvar-Aghili-2005,Jean-Fu-1993,Aghili-2012a,Namvar-Aghili-2004a}, robotic hands
for grasping objects
\cite{Aghili-2010,Cole-Hauser-Sastry-1989,Aghili-2011k,Zheng-Nakashima-Yoshikawa-2000},
parallel manipulators
\cite{Aghili-2010i,Murray-Lovell-1989,Aghili-2019a,Nakamura-Ghodoussi-1989}, impact in robotics \cite{Aghili-2019b},  humanoid robot
and walking robots
\cite{Perrin-Chevallereau-Verdier-1997,Surla-Rackovic-1996,Aghili-2017a}, and
VR/Haptic applications~\cite{Bicchi-Pallottino-Bray-Perdomi-2001,Aghili-2019c}.
Simulation and control of such systems call for corresponding direct
dynamics and inverse-dynamics models, respectively. Mathematically,
constrained mechanical systems are modeled by a set of $n$
differential equations coupled with a set of $m$ algebraic
equations, i.e. Differential Algebraic Equations (DAE). Although
computing the dynamics model is of interest for both simulation and
control, the research done in these two areas are rather divided.
Surveys of the existing techniques for solving DAE may be found in
\cite{GarciadeJalon-Bayo-1994,deJalon-Bayo-1989,Ascher-Petzold-1998,
Gear-1971,Baumgarte-1972,Bicchi-Pallottino-Bray-Perdomi-2001}, while
model-based control of constrained manipulators can be found in
\cite{Raibert-Craig-1981,Khatib-1987,Aghili-2015c,Yoshikawa-1987,Aghili-2010h,Hogan-1985,
West-Asada-1985,Aghili-2010c,Chiaverini-Sciavicco-1993,
Canudas-Siciliano-Bastin-book-1996,Aghili-2011b}.

The classical method to deal with DAE is to express the constraint
condition at the acceleration level. This allows replacement of the
original DAE system with an ODE system by augmenting the inertia
matrix with the second derivative of the constraint equation.
However, this method performs poorly in the vicinity of
singularities
\cite{Chang-Nikravesh-1985,Lin-Hon-1998,Yoon-Howe-1995}, because the
augmented inertia matrix is invertible only with a full rank
Jacobian matrix.

Other methods are based on coordinate partitioning
\cite{Wehage-Haug-1982,Nikravesh-Chung-1982,Liang-Lance-1987} by
using the fact that the $n$ coordinates are not independent because
of the $m$ constraint equations. The motion of the system can be
described by the independent coordinates which can be separated
using an annihilator operator. Although this method may
significantly reduce the number of equations, finding the
annihilator operator is a complex task
\cite{Bicchi-Pallottino-Bray-Perdomi-2001}. Moreover; the sets of
independent and dependent coordinates should be determined first.
But a fixed set of independent coordinates occasionally leads to
ill-conditioned matrices
\cite{GarciadeJalon-Bayo-1994,Blajer-Schiehlen-Schirm-1994} when the
system changes its topology or the number of degrees of freedom. The
concept of coordinate separation is used in
\cite{McClamorch-Wang-1988}for controlling manipulator robots with
constrained end-effectors. The augmented Lagrangian formulation
proposed in
\cite{Bayo-Jalon-1988,Bayo-Ledesma-1996,Cuadrado-Cardenal-Bayo-1997}
can handle redundant constraints and singular situations. However,
this formulation solves the equations of motion through an iterative
process. Nakamura et al. \cite{Nakamura-Yamane-2000} developed a
general algorithm that provides a way to partition the coordinates
into independent and dependent ones even around the singular
configuration, which is suitable for simulation of mechanical
systems with structure-varying kinematic chains. This is a special
case of the projection method proposed herein that allows  generic
constraints which cannot be handled by coordinate partitioning.

There are also efficient algorithms for solving direct dynamics of
constrained systems that are suitable for parallel processing.
Feathersone's work in
\cite{Featherstone-1999,Featherstone-Orin-2000} presents a recursive
algorithm, which is called Divide-and-Conquer (DAC), for calculating
the forward dynamics of general {\em rigid-body} system on a
parallel computer. The central formula for these DAC algorithm takes
the equations of motions of two independent subassembly (rigid-body)
and also a description of how they are to be connected, and the
output is the equation of motion of the assembly, i.e., those of two
articulated-body. Since the equation of acceleration of the assembly
is written in terms of two independent equation of motions, the
formulation is suitable for parallel processing and one can apply
the formula recursively to construct the articulated-body equations
of motions of an entire rigid-body assembly from those of its
constituent parts. The author claims that the DAC algorithm is
computationally effective if a large number of processors, more than
$100$, is available.

Another group of researchers
\cite{Bayo-Ledesma-1996,Park-Chiou-1988,Blajer-1997,Yoon-Howe-1995,Blajer-2002,Aghili-2015b}
focused on other techniques to deal with the problem of accurately
maintaining the constraint condition. Blajer
\cite{Blajer-1997,Blajer-2002} proposed an elegant geometric
interpretation of constrained mechanical systems. Then the analysis
was extended and modified in \cite{Guanfeng-Zexiang-2002} for
control application. The augmented Lagrangian formulation proposed
in \cite{Bayo-Jalon-1988,Bayo-Ledesma-1996,Cuadrado-Cardenal-Bayo-1997}
can handle redundant constraints and singular situations. However,
this formulation solves the equations of motion through an iterative
process.

In the realm of control of constrained multibody system, the vast
majority of the literature is devoted to control of manipulators
with constrained end-effectors. The hybrid position/force control
concept was originally introduced in \cite{Raibert-Craig-1981}, and
then the manipulator dynamic model was explicitly included in the
control law in \cite{Khatib-1987}. The constrained task formulation
with inverse-dynamics controller is developed in
\cite{Yoshikawa-1987,Yoshikawa-Sugie-Tanaka-1988} by assuming that
the Cartesian constraints are linearly independent. Hybrid
motion/force control proposed in
\cite{Luca-Manes-1994,Doty-Melchiorri-1993,Schutter-Bruyinckx-1996}
achieves a complete decoupling between channels of acceleration and
force. In these approaches all joints are assumed to have an
actuator and no redundancy was considered in the kinematic
constraint.

In this work, we propose a new formulation for the direct and the
inverse-dynamics of constrained mechanical systems based on the
notion of a projection operator \cite{Aghili-2005,Aghili-2003}. First,
constraint reaction forces are eliminated by projecting the initial
dynamic equations into the tangent space with respect to the
constraint manifold. Subsequently, the direct dynamics, or the
equations of motion, is derived in a compact form that relates
explicitly the generalized force to the acceleration by introducing
a {\em constraint inertia matrix}, which turns out to be always
invertible. The constraint reactions can then be retrieved from the
dynamics projection in the normal space \cite{Aghili-2005}. Unlike in the other
formulations, the projection matrix is a square matrix of order
equal to the number of dependent coordinates. Since the formulation
of the projector operators is based on pseudo-inverting the
constraint Jacobian (the process not conditioned upon the maximal
rank of the Jacobian), the present approach is valid also for
mechanical systems with redundant constraints and/or singular
configurations, which is unattainable with many other classical
methods. A projected inverse-dynamics control scheme is developed
based on the dynamics formulation. The motion control proves to be
stable while minimizing the weighted Euclidean norm of actuation
force. The notion of the projected inverse-dynamics is further
developed for control of constrained mechanical systems which have
passive joints, i.e., joints with no actuator. This result is
particulary important for control of parallel manipulators. Finally,
a hybrid force/motion control scheme based on the proposed
formulation is presented. Also some useful insights are gained from
the dynamics formulation. For instance, the condition on the inertia
matrix for achieving a complete decoupling between force and motion
equations is rigorously derived.

This paper is organized as follows: We begin with the notion of {\em
linear operator equations} in Section \ref{sec_Linear} by reviewing
some basic definitions and elementary concepts which will be used in
the rest of the paper. Using the projection operator, we derive
models of inverse and  direct dynamics in sections \ref{sec_Inverse}
and  \ref{sec_Direct} which are used as a basis for developing
strategies for simulation and control of constrained mechanical
systems in sections \ref{sec_Simulation} and \ref{sec_Control}.
Section \ref{sec_DiffUnit} presents change of coordinate if there is
inhomogeneity in the spaces of the force and velocity. In section
\ref{sec_Passive}, the inverse-dynamics control scheme is extended
for constrained systems which have some joints with no actuator
(passive joints).

\section{Linear Operator Equations}
\label{sec_Linear}

For any linear operator transformation $ A: \mathbb{R}^{n}
\rightarrow \mathbb{R}^{m}$, range space and null space are
defined as ${\cal R}(A) = \{y \in \mathbb{R}^{m}: \exists x \in
\mathbb{R}^{n} \ni y = A x \}$ and ${\cal N}(A)=\{x \in
\mathbb{R}^{n}: A x =0 \}$, respectively. The linear
transformation maps vector space ${\cal X}$ into vector space
${\cal Y}$. Assume that the Euclidean inner-product is defined in
${\cal X}$, that is elements of vectors, such as $x_1$ and $x_2$,
of ${\cal X}$ have homogeneous units. Then, by definition the
vectors are orthogonal iff their inner-product is zero, i.e.
\begin{equation}\label{eq_dotp}
< x_1 , x_2> = x_1^T x_2 = x_2 ^T x_1 =0
\end{equation}
where the superscript $^T$ denotes transpose. It follows that the
orthogonal complement of any set ${\cal S}$, denoted by ${\cal
  S}^{\perp}$, is the set of vectors each of which is orthogonal to
every vector in ${\cal S}$.

\begin{theorem} \label{RangeNull}
  \cite{Desoer-1970,Golub-VanLoan-1996} The fundamental relationships
  between the range space and the null space associated with a linear
  operator and its transpose are
\begin{equation} \label{eq_rangA}
{\cal R}(A^T)^{\perp} = {\cal N}(A),
\end{equation}
\begin{equation} \label{eq_rangAt}
{\cal R}(A) = {\cal N}(A^T)^{\perp}.
\end{equation}
\end{theorem}
See Appendix \ref{apx_RangeNull} for a proof.

As will be seen in the following sections, it is desirable to be
able to project any vector in $\mathbb{R}^{n}$ to the null space of
$A$ by a projector operator. Let $P \in \mathbb{R}^{n \times n}$ be
the orthogonal projection onto the null space, i.e., ${\cal R}(P) =
{\cal N}(A)$. Note that every orthogonal projection operator has
these properties: $P^2=P$ and $P^T=P$ \cite{Golub-VanLoan-1996}.

The projection operator can be calculated by the {\em singular value
decomposition} (SVD) method
\cite{Golub-VanLoan-1996,Press-Flannery-1988,Belta-Kumar-2002}.
Assuming
\[ r=\mbox{rank}(A), \]
then there exist unitary matrices $U=[U_1 \;\; U_2]$ and $V=[V_1
\;\; V_2]$ (i.e. $U^T U = I$ and $V^T V = I$) so that
\[ A = \left[ \begin{array}{cc}  U_1 & U_2 \end{array} \right]
\left[ \begin{array}{cc} \Sigma & 0 \\ 0 & 0 \end{array} \right]
\left[ \begin{array}{c} V_1^T \\ V_2^T \end{array} \right]
\]
where $\Sigma= \mbox{diag}(\sigma_1 , \cdots, \sigma_r)$, and
$\sigma_1 \geq \cdots \geq \sigma_r \geq0$ are the singular
values. The proof of this statement is straightforward and can be
found, for example, in
\cite{Golub-VanLoan-1996,Press-Flannery-1988}. Since ${\cal N}(A)
= \mbox{span}(V_2)$ \cite{Golub-VanLoan-1996,Press-Flannery-1988},
the projection operator can be calculated by

\begin{equation} \label{eq_VV}
P = V_2 V_2^T
\end{equation}

\subsection{Orthogonal Decomposition and Norm}
From the definition, one can show that projector operator $(I - P)$
projects onto the null space orthogonal ${\cal N}(A)^{\perp}$. Let's
assume that the elements of a vector $x \in \mathbb{R}^{n}$ have
homogeneous measure units, then the vector has a unique orthogonal
decomposition
\[ x = x_{\parallel} \oplus x_{\perp}, \]
where $x_{\parallel} \in {\cal N}(A)$ and $x_{\perp} \in {\cal
N}(A)^{\perp}$. The components of the decomposition can be
obtained uniquely by using the projection operator as
\begin{equation} \label{eq_dcomp}
x_{\parallel} := P x \;\;\;\ \mbox{and} \;\;\;\; x_{\perp} := (I -
P) x.
\end{equation}
The Euclidean norm is defined as
\begin{equation} \label{eq_norm}
\| x \| := <x , x >^{1/2} = (x^T x)^{1/2}
\end{equation}
\begin{remark}
From orthogonality of the subspaces, i.e. $<x_{\parallel},
x_{\perp}>=0$, we can say
\begin{equation} \label{eq_tri_norm}
\| x \|^2 = \| x_{\parallel} \|^2 + \| x_{\perp} \|^2
\end{equation}
\end{remark}
Equation (\ref{eq_tri_norm}) forms the basis for finding an
optimal solution.

\subsubsection{Metric tensor} \label{sec_metric}
The Euclidean inner product and hence the Euclidean norm defined in
(\ref{eq_norm}) are non-invariant quantities, if there is
inhomogeneity in the units of the elements of vector $x$. With the
same token, the projection matrix (\ref{eq_dcomp}) and the
decomposition are not invariant and hence the minimum-norm solution
may depend on the measure units chosen. This is because components
with different units are added together in (\ref{eq_dcomp}).

To circumvent the quandary of the measure units, we consider the
following transformation
\begin{equation} \label{eq_xW}
x_W := W^{1/2} x,
\end{equation}
and assume that the vector $x_W$ has components with the same
physical units. Then, a physically consistent Euclidean inner
product and Euclidean norm exists on the new space
\cite{Doty-Melchiorri-1993}, i.e.
\begin{equation} \label{eq_normW1}
\| x_W \| = < x_W , x_W >^{1/2} = <x , W x>^{1/2}
\end{equation}
The symmetric, positive definite matrix $W$ is called a {\em
metric tensor} of the n-space. Note that the Euclidean norm of the
new coordinate is tantamount to the weighted-norm, that is $\| x
\|_W = \| x_W \|$, where
\begin{equation} \label{eq_normW}
\| x \|_W = (x^T W x)^{1/2}.
\end{equation}

Furthermore, denoting $A_W :=AW^{1/2}$, one can say $A_W x_W=0$. Let
$P_W$ be the projection operator onto the null space of $A_W$. Then,
mapping $P_W$ is dimensionless and invariant.

\section{Decomposition of the Acceleration}
\label{sec_decomposition}

The kinematics of a constrained mechanical system can be represented
by a set of $m$ nonlinear equations ${\Phi}(q) =[\phi_1(q), \cdots,
\phi_m(q)]^T = 0$, where $q \in \mathbb{R}^{n}$ is the vector of
generalized coordinate, and $m \leq n$. Without loss of generality,
we consider time-invariant (scleronomic) constraint conditions, but
the methodology can be readily extended to a time varying case
(rheonomic). By differentiating the constraint equation with respect
to time, we have
\begin{equation} \label{eq_dq}
A \dot{q}= 0
\end{equation}
where $A = {\partial \Phi }/{\partial q}$ is the Jacobian of the
constraint equation with respect to the generalized coordinate. For
brevity of notation, in the following, we assume that the elements
of the force and velocity vectors have homogeneous units. This
assumption will be relaxed in Section \ref{sec_DiffUnit} by changing
of the coordinates  similar to (\ref{eq_xW}).

Equation (\ref{eq_dq}) is expressed in form of the linear operation
equation. This matrix equation specifies that any admissible
velocity must belong to the null space of the Jacobian matrix, that
is, $\dot{q} \in {\cal N}(A)$. Thus, the constraint equation
(\ref{eq_dq}) can be expressed by the notion of the projection
operator, i.e.,
\begin{equation} \label{eq_(I-P)dq}
\dot{q}_{\perp} \equiv (I - P) \dot{q} =0,
\end{equation}
Time differentiation of the above equation yields
\begin{equation} \label{eq_(I-P)ddq}
\ddot{q}_{\perp} \equiv (I - P) \ddot{q} = C \dot{q},
\end{equation}
where $C := \frac{d}{dt} P$ which, in turn, can be obtained from
(\ref{eq_VV}) by
\begin{equation} \label{eq_C}
C = S + S^T \;\;\;\; \mbox{where} \;\;\; S= \dot{V}_2 V_2^T
\end{equation}

It is apparent from equation (\ref{eq_(I-P)ddq}) and
(\ref{eq_(I-P)dq}) that, unlike the case of velocity, the null
space orthogonal component of the acceleration is not always zero
-- a physical interpretation of (\ref{eq_(I-P)ddq}) is given in
Section \ref{sec_example}. Equation (\ref{eq_(I-P)ddq}) expresses
the component of acceleration produced exclusively by the
constraint and not by dynamics. As will be seen in Section
\ref{sec_Direct}, this equation can compliment the dynamics
equation in order to provide sufficient independent equations for
solving the acceleration.

\subsection{Calculating $P$ based on Pseudo-Inversion}

Many mature algorithms and numerical techniques are available for
computing the {\em pseudo-inverse}
\cite{Press-Flannery-1988,Golub-VanLoan-1996,BenIsrael-1980}. There
are also computer programs that can solve SVD and pseudo-inverse in
real-time and non real-time, for instance DSP Blockset of Matlab
\cite{Matlab_DSP}. Therefore, it may be useful to calculate the
matrices $P$ and $C$ based on pseudo-inversion.

Let $A^+$ denote the pseudo-inverse of $A$. Then, the projection
operator can be calculated by
\begin{equation} \label{eq_PA+}
P= I - A^+ A.
\end{equation}
Also, one can obtain matrix $C$ through the pseudo-inversion as
follows. Differentiation of (\ref{eq_dq}) with respect to time
leads to
\[ A \ddot{q}= - \dot{A} \dot{q} \]
The theory of linear systems of equations
\cite{Golub-VanLoan-1996,BenIsrael-1980,Nakamura-1991} establishes
that the particular solution, i.e., the ${\cal N}^{\perp}$ component
of the acceleration, can be obtained from the above equation as
$\ddot{q}_{\perp}= - A^+ \dot{A} \dot{q}$. Hence
\begin{equation}
C \dot q = -A^+ \dot{A} \dot q
\end{equation}

Assuming the elements of generalized coordinate have identical
units, then matrices $P$ and $C$ have homogeneous units, i.e. $P$ is
dimensionless and the dimension of $C$ is $s^{-1}$. Therefore, $P$
and $C$ are invariant under unit changes.

Note that the inconsistency problem which may arise in computing the
pseudo-inverse because of existence  of the components of different
units can be solved by including the metric of the n-space in
computing the pseudo-inverse \cite{Blajer-1997,Blajer-2001a}.

\section{Projected Inverse-Dynamics}
\label{sec_Inverse}
Consider a constrained mechanical system with Lagrangian ${\cal L}=
{\cal T}- {\cal V}$, where ${\cal T}=\frac{1}{2}\dot{q}^T M \dot{q}$
and ${\cal V}(q)$ are the kinetic and the potential energy
functions, and $M \in \mathbb{R}^{n \times n}$ is the inertia
matrix. The fundamental equation of differential variational
principles of a mechanical system containing constraint can be
written as \cite{deJalon-Bayo-1989}
\begin{equation} \label{eq_deltaqf}
\delta q^T f_{tot} = 0,
\end{equation}
where $f_{tot}= \varphi - (f - {\cal F})$,
$\varphi=\frac{d}{dt}\left(\frac{\partial {\cal L}}{\partial
\dot{q}}\right) - \frac{\partial {\cal L}}{\partial q}$, ${f} \in
\mathbb{R}^{n}$ is the vector of generalized input force, and ${\cal
F}\in \mathbb{R}^{n}$ is the {\em generalized constraint force},
which is related to the {\em Lagrange multipliers} ${\lambda} \in
\mathbb{R}^{m}$ by
\begin{equation} \label{eq_F}
{\cal F} = A^T {\lambda} \in {\cal R}(A^T).
\end{equation}
Then, the equations describing the system dynamics can be obtained
as
\begin{equation} \label{eq_Mddq}
M \ddot{q} + h(q, \dot{q}) = f - {\cal F}
\end{equation}
\begin{equation} \label{eq_constraint}
\Phi (q) = 0,
\end{equation}
where vector $h(q, \dot{q}) \in \mathbb{R}^{n}$ contains the
Coriolis, centrifugal, gravitational terms. In solving the DAE
equations (\ref{eq_Mddq})-(\ref{eq_constraint}), it is typically
assumed that: (i) the inertia matrix is positive definite and hence
invertible (ii) the constraint equations are independent, i.e. the
Jacobian matrix is not rank-deficient
\cite{GarciadeJalon-Bayo-1994,deJalon-Bayo-1989,Ascher-Petzold-1998,
Gear-1971,Baumgarte-1972,Bicchi-Pallottino-Bray-Perdomi-2001}. In
this work, we solve the equations without relying on the second
assumption.

From (\ref{eq_F}) and by virtue of Theorem \ref{RangeNull}, one can
immediately conclude that ${\cal F} \in {\cal N}(A)^{\perp}$. In
other words, the projection operator $P$ is an annihilator for the
constraint force, i.e. $P {\cal F}=0$. Therefore, the constraint
force can be readily eliminated from equation (\ref{eq_Mddq}) if the
equation is projected on $P$, i.e.
\begin{equation} \label{eq_PMddq}
P M \ddot{q} = P (f - h) .
\end{equation}
Equation (\ref{eq_PMddq}) is called {\em projected inverse-dynamics}
of a constrained multibody system that is expressed in the so-called
{\em descriptive form}. This is because matrix $P M$ is singular and
hence the acceleration cannot be computed from the equation through
matrix inversion.

\section{Direct Dynamics}
\label{sec_Direct}

As mentioned earlier, the acceleration cannot be determined uniquely
from equation (\ref{eq_PMddq}), because there are fewer independent
equations than unknowns. Nevertheless; equations (\ref{eq_(I-P)ddq})
and (\ref{eq_PMddq}) are in orthogonal spaces and thus cannot cancel
out each other. Therefore, a unique solution can be obtained by
solving these two equations together. To this end, we simply
multiply equation (\ref{eq_(I-P)ddq}) by $M$ and then add both sides
of the equation to those of (\ref{eq_PMddq}). After factorization,
the resultant equation can be written concisely in the following
form;
\begin{equation} \label{eq_calMddq}
M_c \ddot{q} = P(f - h) + C_c \dot{q}
\end{equation}
where $C_c := M C$, and $M_c \in \mathbb{R}^{n \times n}$ is called
{\em constraint inertia matrix} which is related to the
unconstrained inertia matrix $M$---assuming a symmetric inertia
matrix---by
\begin{equation} \label{eq_Mc}
M_c := M + \tilde{M},
\end{equation}
and
\begin{equation} \label{eq_tM}
\tilde{M} := P M  - (P M)^T.
\end{equation}
Equation (\ref{eq_calMddq}) constitutes the so-called {\em direct
dynamics} of a constrained multibody system from which the
acceleration can be solved. It is worth mentioning that if $M$
commutes with $P$, then $\tilde{M}=0$ and hence $M_c= M$. To compute
the acceleration from (\ref{eq_calMddq}) requires that the
constraint inertia matrix be invertible.

\begin{theorem}
\label{ConstraintInertia} If the unconstrained inertia matrix $M$ is
positive-definite (p.d.), then the constraint inertia matrix $M_c$
is p.d. too.
\end{theorem}
{\sc Proof:} It is evident from (\ref{eq_tM}) that $\tilde{M}$ is a
skew-symmetric matrix, i.e. $\tilde{M}^T =- \tilde{M}$.
Consequently, adding $\tilde{M}$ to the inertia matrix in equation
(\ref{eq_Mc}) preserves the positive definiteness property of the
inertia matrix. This is because, for any vector $z \in
\mathbb{R}^{n}$ we can say
\[ z^T \tilde{M} z = 0. \]
Therefore, one can conclude that $z^T {M} z = z^T M_c z$, or
\[ M \;\;\; \mbox{is p.d.} \;\;\; \Longleftrightarrow \;\;\; M_c
\;\;\;\;   \mbox{is p.d.}. \]
$\Box$ .

Theorem \ref{ConstraintInertia} is pivotal in showing the usefulness
of the dynamics equation (\ref{eq_calMddq}); it signifies that the
constraint inertia matrix is always invertible regardless of the
constraint condition. Therefore, the acceleration can be always
obtained from (\ref{eq_calMddq}).

\begin{remark}
\label{ForceContribution}
  Equation (\ref{eq_calMddq}) signifies that only the
  null space component of the generalized input force contributes to the
  motion of a constrained mechanical system, as the projector in the right-hand-side
  (RHS) of the equation filters out all forces lying in the null space orthogonal. This fact
  is exploited in Section \ref{sec_MotionCntr} for an optimal control scheme.
\end{remark}
It can be envisaged from Remark \ref{ForceContribution} that it is
useful to decompose the generalized input force into two orthogonal
components
\[ f = f_{\parallel} \oplus f_{\perp}, \]
where $f_{\parallel} \in {\cal N}$ and $f_{\perp} \in {\cal
  N}^{\perp}$ are called {\em acting input force} (potent) and {\em
  passive input force} (impotent), respectively. The decomposition of
the generalized input force can be carried out by the projection
operator according to (\ref{eq_dcomp}). Now, the equation of motion
can be written as
\begin{equation} \label{eq_InvDyn}
\ddot{q} = M_c^{-1} (f_{\parallel} -h_{\parallel} + C_c \dot{q})
\end{equation}
where the nonlinear vector is decomposed in the same way as the
generalized force, i.e., $h=h_{\parallel}+h_{\perp}$. Equation
(\ref{eq_InvDyn}) is the so-called {\em equation of motion} of a
constrained mechanical system in a compact form. It is worth
mentioning that only one matrix inversion operation is required in
(\ref{eq_InvDyn}) which is one less than in the standard Lagrangian
method (see Appendix \ref{apx_Lagrang}).

\begin{remark} \label{Kinetic}
  Since $\tilde{M}$ does not produce any kinetic energy, the total kinetic energy
  associated with the constrained system is
  ${\cal T}= \frac{1}{2} \dot{q}^T M_c \dot{q} = \frac{1}{2} \dot{q}^T M
  \dot{q}$.
\end{remark}

\subsection{Constraint Inertia Matrix}
The constraint inertia matrix doesn't have a unique definition
because there are many ways that equations (\ref{eq_PMddq}) and
(\ref{eq_(I-P)ddq}) can be combined together. Although all dynamics
formulations thus obtained are equivalent, each one may have a
ceratin computational advantage over the others.

\subsubsection{Symmetric Inertia Matrix}
$M_c$ is a p.d. matrix but not a symmetric one. In the following, we
present an alternative dynamics formulation in which the inertia
matrix appears both p.d. and symmetric. Equation (\ref{eq_(I-P)ddq})
together with the decomposition of the acceleration imply that
$\ddot q = P \ddot q + C \dot q$, which can be substituted in
(\ref{eq_PMddq}) to give
\begin{equation} \label{eq_PMP}
PMP \ddot q = P(f-h) - PC_c \dot q.
\end{equation}
Now premultiply equation (\ref{eq_(I-P)ddq}) by $(I-P)M$ and then
add both sides of the equation thus obtained with those of
(\ref{eq_PMP}) yields
\begin{equation} \label{eq_M'cddq}
M'_c \ddot q = P(f-h) + C'_c \dot q,
\end{equation}
where
\begin{equation} \label{M'c}
 M'_c := PMP + (I-P) M (I-P),
\end{equation}
and
\[ C'_c := (I-2P) C_c.\]
It is worth mentioning that $I-2P$ represents a {\em reflection
operator}.
\begin{proposition}
If matrix $M$ is symmetric and p.d., then $M'_c$ is symmetric and
p.d. too.
\end{proposition}
{\sc Proof:} It is apparent from (\ref{M'c}) that $M'_c$ is a symmetric matrix .
Moreover, positive definiteness of matrix $M'_c$ can be shown by an
argument similar to the previous case. Again for any vector $z \neq
0 \in \mathbb{R}^{n}$ and from definition (\ref{M'c}), we can say
$z^TM'_c z=z_{\parallel}^T M z_{\parallel} +z_{\perp}^T M
z_{\perp}$, where $z_{\parallel} \in {\cal N}$ and $z_{\perp} \in
{\cal N}^{\perp}$, and $z_{\parallel}^T M z_{\parallel} \geq 0$ and
$z_{\perp}^T M z_{\perp} \geq0$. Both decomposed components of the
non-zero vector $z$ cannot be zero, i.e., $z_{\parallel}=0
\Rightarrow z_{\perp} \neq0$ and vice versa. Therefore, only one of
the quadratic functions can be zero and that implies their summation
is non-zero and positive. Thus $M'_c$ is a p.d. matrix $\Box$.

\subsubsection{Parameterized Inertia matrix}

Alternatively, the constraint inertia matrix can be parameterized in
terms of an arbitrary scalar. To this end, let's first premultiply
equation (\ref{eq_(I-P)ddq}) by a scalar $\gamma$ and then add both
sides of the resultant equation  with those of (\ref{eq_PMddq}).
That gives the standard dynamics formulation similar to
(\ref{eq_M'cddq}) with the following parameters
\begin{equation} \label{eq_M''c}
M''_c := P M + \gamma(I - P),
\end{equation}
and
\[C''_c :=\gamma C. \]

\begin{proposition}
If $M$ is an invertible matrix, then $M''_c$ is always invertible
too.
\end{proposition}
{\sc Proof:} In a proof by contradiction we show that $M''_c$ should be a
full-rank matrix. If matrix $M''_c$ is not of full-rank, then there
must exist at least one non-zero vector $\xi \neq0$ lying in the
matrix null space, that is $M''_c \xi =0$, or
\begin{equation} \label{eq_Mxi}
P M  \xi + \gamma(I - P) \xi =0
\end{equation}
The two terms of the above equation are in two orthogonal subspaces
and cannot cancel out each other. Hence, in order to satisfy the
equation, both terms must be identically zero, i.e. $P M \xi=0$ and
$\gamma(I - P) \xi=0$. The former and the latter equations imply
that $y=M \xi \in {\cal N}^{\perp}$ and $\xi \in {\cal N}$,
respectively. Therefore, one can conclude that $y$ is perpendicular
to vector $\xi$, i.e. $\xi^T y =0$, or that
\begin{equation} \label{eq_xiMxi}
\xi^T M \xi =0,
\end{equation}
which is a contradiction because $M$ is a positive-definite matrix.
Therefore, the null set of $M''_c$ is empty and the matrix is always
invertible, and this completes the proof $\Box$.

Since $P$ is dimensionless, the scalar $\gamma$ has the dimensions
of mass. Therefore, the value of $\gamma$ should be comparable to
that of $M$ to avoid any numerical pitfall in the matrix inversion
--- a logical choice is $\gamma=\| M \|$; yet, certain $\gamma$ may
lead to the minimum condition number of $M_c$, which is desired for
matrix inversion.

\subsubsection{A Comparison of different Constraint Inertia Matrices}
Theoretically, all the inverse-dynamics formulations presented here
are equivalent, and they should yield the same result. However, from
a numerical point of view, each has a certain advantage over the
others that can lead to simplification of simulation or control. In
summary,
\begin{itemize}
\item $M_c$ is a p.d. matrix but not a symmetric one. If $P$ commutes with $M$, then
$M_c=M$.
\item $M'_c$ is a symmetric and p.d. matrix; hence it
physically exhibits the characteristic of an inertia matrix.
However, computing $M'_c$ involves three additional matrix
multiplication operations compared to $M_c$.
\item $M''_c$ is an invertible matrix, but it is neither p.d. nor symmetric. Nevertheless,
computing of $M''_c$ requires less computation effort compared to
the others.
\end{itemize}

\subsection{Constraint Force and Lagrange Multipliers}

Equation (\ref{eq_InvDyn}) expresses the generalized acceleration of
a constrained multibody system in a compact from without any need
for computing the Lagrange multipliers. Yet, in the following we
will retrieve the constraint force by projecting equation
(\ref{eq_Mddq}) onto $I - P$, that gives
\[ {\cal F} =(f_{\perp} - h_{\perp}) - (I - P) M \ddot{q}. \]
Now, substituting the acceleration from (\ref{eq_InvDyn}) into the
above equation gives
\begin{equation} \label{eq_lamb}
{\cal F}= (f_{\perp} - h_{\perp}) - \mu(f_{\parallel} -
h_{\parallel}) - \mu C_c \dot{q},
\end{equation}
where $\mu = (I - P) \alpha$, and $\alpha = M M_c^{-1}$ is the
ratio of the two inertia matrices.

Equation (\ref{eq_lamb}) implies that the constraint force can
always be obtained uniquely, but this may not be true for the
Lagrange multipliers. Having calculated the constraint force from
(\ref{eq_lamb}), one may obtain the Lagrange multipliers from
equation (\ref{eq_F}) through pseudo-inversion, i.e., $\lambda =
A^{+T} {\cal F} + \lambda_n$ where $\lambda_n \in {\cal N}(A^T)$ is
the homogeneous solution. By virtue of Theorem \ref{RangeNull}, we
can also say that $\lambda_n \in {\cal R}(A)^{\perp}$. This, in
turn, implies that $\lambda_n$ is a non-zero vector only if the
Jacobian matrix is rank deficient, i.e. $r < m$---recall that
$r=\mbox{rank}(A)$.
\begin{remark} \label{UniqueLambda}
  The vector of Lagrange multipliers can be determined uniquely iff
  the Jacobian matrix is full-rank. In that case, there is a
  one-to-one correspondence between ${\cal F}$ and ${\lambda}$.
  Otherwise, the ${\cal R}(A)^{\perp}$ component of the Lagrange
  multiplies is indeterminate.
\end{remark}

\subsection{Decoupling}

\begin{figure}[tbhp]
\centering{\includegraphics[width=7cm]{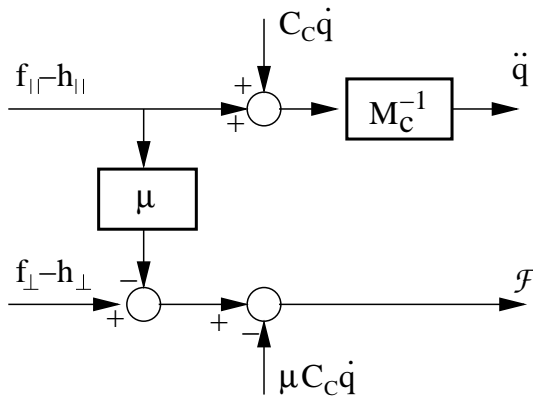}}
\caption{The input/output realization of a constrained mechanical system based on
decomposition of the generalized input force.}
\label{fig_constrained}
\end{figure}

Fig.\ref{fig_constrained} illustrates the input/output realization
of a constrained mechanical system based on (\ref{eq_InvDyn}) and
(\ref{eq_lamb}). The input channels $f_{\parallel}$ and $f_{\perp}$
are the potent and the impotent components of the generalized input
force, while the output channels are the acceleration and the
constraint force, $\ddot{q}$ and ${\cal F}$, respectively. It is
apparent form the figure that the acceleration is only affected by
$f_{\parallel}$ and not by $f_{\perp}$ whatsoever. However, the
constraint force output, in general, can be affected by two inputs:
by $f_{\perp}$ directly, and by $f_{\parallel}$ through the
cross-coupling channel $\mu$. The cross-coupling channel is disabled
if the inertia matrix satisfies a certain condition which is stated
in Proposition \ref{th_Decoupling}.

\begin{proposition} \label{th_Decoupling}
  The equations of the constraint force and the acceleration are
  completely decoupled, i.e. the cross coupling $\mu$ vanishes if
  the null space of the constraint Jacobian is invariant under $M$. That
  is, the inertia matrix should have this property: $\{ \forall x \in
  {\cal N}(A) \; : \; M x \in {\cal N}(A) \}$.
\end{proposition}
The proof is given in Appendix \ref{apx_decoupling}.

Mechanical systems satisfying the condition in Proposition
\ref{th_Decoupling} are called {\em decoupled constrained mechanical
systems}. The equation of constraint force of such a system is
reduced to
\[ {\cal F}= (f_{\perp} - h_{\perp})  - \mu C_c \dot{q}. \]
In that case, the constraint force is determined exclusively by
the passive input force that leads to a simple force control
scheme, as will be seen in Section \ref{sec_ForceCntr}.

\subsection{Illustrative Example and Geometrical Interpretation}
\label{sec_example}
\begin{figure}[tbhp]
\centering{\includegraphics[width=7cm]{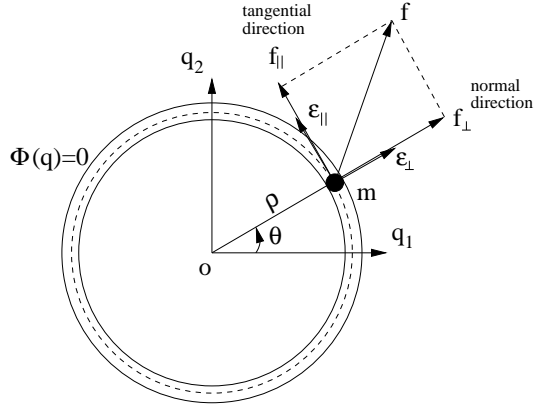}}
\caption{An illustrative example.} \label{fig_circle}
\end{figure}

A particle of mass $m$ moves on a circle of radius $\rho$; see
Fig.\ref{fig_circle}. Assume $q=[q_1 \; q_2]^T$ to be the
generalized coordinate. The constraint equation is
\[ \Phi(q)=(q_1^2 + q_2^2)^{1/2} - \rho=0 \]
which yields the Jacobian and its time-derivative as $A=[q_1/\rho
\; q_2/\rho]$ and $\dot{A}=[\dot{q}_1/\rho \; \dot{q}_2/\rho]$,
and the pseudo-inverse is $A^+=[q_1/ \rho \; q_2/ \rho]^T$. Then,
from (\ref{eq_PA+}) we have
\begin{equation} \label{eq_Pm}
P=\frac{1}{\rho^2} \left[\begin{array}{cc} q_2^2 & -q_1 q_2
\\ -q_1 q_2 & q_1^2 \end{array} \right] = \left[\begin{array}{cc}
\sin^2 \theta & -\sin \theta \cos \theta \\ -\sin \theta \cos
\theta & \cos^2 \theta
\end{array} \right]
\end{equation}
Let unit vectors $\epsilon_{\parallel} =[-\sin \theta \; \cos
\theta]^T$ and $\epsilon_{\perp}=[\cos \theta \; \sin \theta]^T$
represent the tangential and normal directions as shown in
Fig.\ref{fig_circle}. Then, from (\ref{eq_Pm}) we have
\[P = \epsilon_{\parallel} \epsilon_{\parallel}^T \;\; , \;\;
I - P = \epsilon_{\perp} \epsilon_{\perp}^T. \] The above equations
represent the geometrical interpretation of the projection
operators. Note that the normal component of the acceleration, $
\ddot{q}_{\perp} = C \dot{q}$, which can be interpreted as the
Centripetal acceleration, is given by
\begin{eqnarray}
C \dot{q} &=& - \frac{1}{\rho^2} \left[ \begin{array}{c}
q_1(\dot{q}_1^2 + \dot{q}_2^2) \\ q_2(\dot{q}_1^2 + \dot{q}_2^2)
\end{array} \right] \\
 &=& - \frac{\| \dot{q} \|^2}{\rho} \epsilon_{\perp} = - \rho \dot{\theta}^2 \epsilon_{\perp}
\end{eqnarray}
Let $f$ denote the external force applied on the particle. Observe
that the decomposition of the force is aligned with the tangential
and the normal directions to the circular trajectory shown  in
Fig.\ref{fig_circle}. Finally, we arrive at the following set of
dynamics equations:
\[m \ddot{q} = f_{\parallel} \]
\[{\cal F}= f_{\perp} + m \frac{\| \dot{q} \|^2}{\rho} \epsilon_{\perp}\]

\section{Simulation of Constrained Multibody Systems}
\label{sec_Simulation}

To simulate the dynamics of a constrained multibody system, one can
make use of the acceleration model in (\ref{eq_InvDyn}). Having
computed the generalized acceleration from the equation, one may
proceed a simulation by integrating the acceleration to obtain the
generalized coordinates. However, the integration inevitably leads
to drift that eventually results in a large constraint error.
Baumgarte's stabilization term \cite{Baumgarte-1972} is introduced
to ensure exponential convergence of the constraint error to zero.
However, this creates a very fast dynamics which tends to slow down
the simulation. In this section, we use the pseudo-inverse for
correcting the generalized coordinate in order to maintain the
constraint condition precisely. It should be noted that using the
pseudo-inverse here does not impose any extra computation burden,
because the pseudo-inverse has to be obtained to compute the
acceleration anyway.

Having obtained the velocity through integrating the acceleration,
one may obtain the generalized coordinate by integration
\begin{equation} \label{eq_init}
\{ q \}_{t}= \{ q \}_{t-\Delta T}  + \int_{t-\Delta T}^t \dot{q} d
\tau,
\end{equation}
where $\Delta T$ depicts integration time step. However, the
constraint condition may be violated slightly because of integration
drift. Let $q_0$ denote the coordinate after a few integration steps
and $\Phi(q_0) \neq0$. Now, we seek a small compensation in the
generalized coordinate $\delta q*= q^* - q_0 $ such that the
constraint condition is satisfied: that is, a set of nonlinear
equations $\Phi(q^*)=0$ must be solved in terms of $q^*$. The
Newton-Raphson method solves a set of nonlinear equations
iteratively based on linearized equations.

The constraint equation can be written by the first order
approximation as
\[ \Phi(q_0+ \delta q ) = \Phi(q_0) + A \delta {q} + {\cal O}(\delta q ^2) =0 \]
Neglecting the ${\cal O}(\delta q ^2)$ term, one can obtain the
solution of the linear system using any generalized inverse of the
Jacobian. The pseudo-inverse yields the minimum-norm solution, i.e.,
$\min_{A \delta {q}=-\Phi(q_0)}\|{\delta q \|}$. Therefore, the
following loop
\begin{equation} \label{eq_Newton}
q_{k+1} = q_k - A^+ {\Phi}(q_k)
\end{equation}
may be worked out iteratively until the error in the constraint
falls into an acceptable tolerance, e.g. $\| {\Phi} \| \leq
\epsilon$.

The condition for local convergence of multi-dimensional
Newton-Raphson (NR) iteration can found, e.g., in
\cite{Kelley-1995,Hokala-Valtonen-2001,GarciadeJalon-Bayo-1994}.
Although it is known that NR iteration will not always converge to a
solution, the convergence is guaranteed if the initial approximation
is close enough to a solution
\cite{Kelley-1995,Hokala-Valtonen-2001}.

\begin{theorem} \label{theorem_NR} \cite{Kelley-1995} Assume that
$\Phi$ is differentiable in an open set $\Omega \subset
\mathbb{R}^{n}$, i.e., the Jacobian matrix $A$ exists, and that $A$
is Lipschitz continuous. Also assume that a solution $q^* \in
\Omega$ exists, and that $A(q^*)$ is nonsingular. Then under these
assumptions, if the initial start point is sufficiently close to the
solution, the convergence is quadratic, that is $\exists K>0$ such
that
\end{theorem}
\[ \| \Phi(q_{k+1}) \| \leq K \| \Phi(q_{k}) \|^2. \]

One can expect that the drift, and hence the inial constraint error,
can be reduced by decreasing the integration time step. It should be
pointed out that the iteration loop (\ref{eq_Newton}) corrects the
error in constraint coordinate caused by the integration process.
Since the drifting error within a single integration time step
$\Delta T$ is quite small, the initial estimate given by
(\ref{eq_init}) cannot be far from the exact solution. Therefore, as
shown by experiments, a fast convergence is achieved even though the
iteration loop (\ref{eq_Newton}) is called once every few time
steps.

Finally, the simulation of a constrained mechanical system based on
the projection method can be proceeded in the following steps:
\begin{enumerate}
\item{compute the acceleration from equation (\ref{eq_InvDyn}).}

\item{obtain the states $\{ q,\dot{q}\}$ as a result of
 numerical integration of the acceleration.}

\item{in the case constraint error exceeds the tolerance, carry
out iteration (\ref{eq_Newton}), upon convergence or counting the
maximum number of iterations go to step(1).}
\end{enumerate}

\section{Control of Constrained Multibody Systems.}
\label{sec_Control}

In this section we discuss the position and/or force control of
constrained multibody systems based on the proposed dynamics
formulation. The I/O realization of a constrained mechanical system
is depicted in Fig.\ref{fig_constrained}, which will be used
subsequently as a basis for development of control algorithm. In
fact, the topology of a control system can be inferred from the
figure by considering the decomposed components of the generalized
input force, $f_{\parallel}$ and $f_{\perp}$, as the corresponding
control inputs for position and force feedback loops.

Due to the decoupled nature of the acceleration channel, an
independent position feedback loop can be applied. The input channel
$f_{\perp}$  is directly transmitted to the constraint force,
$f_{\parallel}$, and the velocity enters as a disturbance and hence
must be compensated for in a feedforward loop. Note that in the case
of decoupled mechanical systems, where the cross-coupling channel
vanishes, $f_{\perp}$ exclusively determines the constraint force.

\subsection{Motion Control Using Projected Inverse-Dynamics Control}
\label{sec_MotionCntr}

Due to  presence of only $r$ independent constraints, the actual
number of degrees of freedom of the system is reduced to $k \leq
n-r$. Thus, in principle, there must be $k$ independent coordinates
$\theta \in \mathbb{R}^{k}$ from which the generalized coordinates
can be derived, i.e., $q= \psi(\theta)$.  Now, differentiation of
the given function with respect to time gives
\begin{equation} \label{eq_dtheta}
\dot{q}= \Lambda \dot{\theta},
\end{equation}
\begin{equation} \label{eq_ddtheta}
\ddot{q}= \Lambda \ddot{\theta} + \dot{\Lambda} \dot{\theta},
\end{equation}
where $\Lambda = \partial \psi / \partial \theta \in \mathbb{R}^{n
\times k}$. Since $\theta(q) = [\theta_1(q), \cdots, \theta_k(q)]^T$
constitutes a set of independent functions, the Jacobian matrix
$\Lambda$ must be of full rank. (The proof is in Appendix
\ref{apx_rank}.). It is also important to note that any admissible
function $\psi(\cdot)$ must satisfy the constraint condition, i.e.,
\[ \Phi(\psi(\theta)) =0  \;\;\;\;\;\;\;\; \forall \theta \in \mathbb{R}^k . \]
Using the chain-rule, one can obtain the time-derivative of the
above equation
\begin{equation} \label{eq_partial}
\frac{\partial \Phi}{\partial q} \; \frac{\partial \psi}{\partial
\theta} \dot{\theta} = A \Lambda \dot{\theta} =0 \;\;\;\;\;\;\;
\forall \dot{\theta} \in \mathbb{R}^k .
\end{equation}
Since $\Lambda$ is a full-rank matrix, the only possibility for
(\ref{eq_partial}) to happen is that
\begin{equation} \label{eq_RangeLamb}
{\cal R}(\Lambda) = {\cal N}(A)
\end{equation}

Substituting the acceleration from (\ref{eq_ddtheta}) into the
inverse-dynamics equation (\ref{eq_PMddq}) gives the dynamics in
terms of the reduced-dimensional coordinate
\begin{equation} \label{eq_PMtet}
P M(\Lambda \ddot{\theta} + \dot{\Lambda} \dot{\theta} ) =
f_{\parallel} - h_{\parallel}
\end{equation}
Let $\{ \theta_d(t) , \dot{\theta}_d(t), \ddot{\theta}_d(t) \}$
denote the desired trajectory of the new coordinates. Now, we
propose the {\em projected inverse-dynamics control} (PIDC) law as
follow
\begin{equation} \label{eq_fs}
f_{\parallel}^c = h_{\parallel} +  P M u_p,
\end{equation}
where $u_p$ is an auxiliary control input as
\begin{equation} \label{eq_up}
u_p = \dot{\Lambda} \dot{\theta} + \Lambda ( \ddot{\theta}_d + G_D
\dot{e}_p + G_P e_p ),
\end{equation}
$e_p={\theta}_d-{\theta}$ is the position tracking error, and
$G_P>0$ and $G_D>0$ are the PD feedback gains. In the following,
superscript $c$ is used to denote control input.

\begin{theorem} \label{optimal}
  While demanding minimum-norm control input, the
  projected inverse-dynamics control law (\ref{eq_fs})-(\ref{eq_up})
  stabilizes the position tracking error, i.e. $\theta(t)
  \rightarrow \theta_d(t) \;\;\; \mbox{as} \;\;\;\; t \rightarrow
  \infty$.
\end{theorem}
{\sc Proof:} Firstly, we prove exponential stability of the position error.
From equations (\ref{eq_PMtet}) -- (\ref{eq_up}), one can conclude
that the proposed control law leads to the following equation for
the tracking error
\begin{equation} \label{eq_PMLe}
P M \Lambda \left[ \ddot{e}_p + G_D \dot{e}_p + G_P e_p \right]
=0.
\end{equation}
To show that the expression within the bracket is zero, we need to
show that the matrix $P M \Lambda$ is full-rank. In the following we
will show that the matrix cannot have any null space and hence is
full rank. If the matrix has a null space, then $\exists x \neq 0 \;
\ni \; P M \Lambda x =0$. Let us define $\xi = \Lambda x$. Recall
that $\Lambda$ is a full-rank matrix and that ${\cal R}(\Lambda) =
{\cal N}(A)$---see (\ref{eq_dtheta}). Hence, $\xi \neq 0 $ and $\xi
\in {\cal N}$. On other hand, $P M \xi =0$ implies that $M \xi \in
{\cal N}(A)^{\perp}$, and hence it is perpendicular to $\xi$, i.e.
$\xi^T M \xi =0$. But, this is a contradiction because $M$ is a p.d.
matrix. Consequently, ${\cal N}( P M \Lambda) = \emptyset $, and it
follows from (\ref{eq_PMLe}) that
\[ \ddot{e}_p + G_D \dot{e}_p + G_P e_p  =0. \]
Hence, the error dynamics can be stabilized by selecting adequate
gains, that is $ \theta \rightarrow \theta_d \;\;\; \mbox{as}
\;\;\;\; t \rightarrow \infty$. Moreover, due to orthogonality of
the decomposed generalized input force, we can say
\[ \| f^c \|^2 = \| f_{\parallel}^c \|^2 + \| f_{\perp}^c \|^2. \]
From the above norm relation, it is clear that $f_{\parallel}^c$ is
the minimum norm solution, since any other solution must have a
component in $f_{\perp}^c$ and this would increase the overall norm.
Therefore, setting $f_{\perp}^c=0$ results in minimum norm of
generalized input force subjected to producing the desired motion,
i.e.,
\begin{equation} \label{eq_opt}
f^c = f_{\parallel}^c \;\;  \Longleftrightarrow  \;\;
\min_{\theta(t) \rightarrow \theta_d(t) }  \| f^c \|.
\end{equation} $\Box$

\subsection{Elements of Generalized Coordinate with Inhomogeneous
Units} \label{sec_DiffUnit}

So far, we have assumed that the elements of the generalized
velocity and the generalized input force have homogeneous units.
Otherwise the minimum-norm solution of the generalized force,
(\ref{eq_opt}), makes no physical sense if the manipulator has both
revolute and prismatic joints. In this section, we assume the vector
of the generalized force to have a combination of force and torque
components, and the vector of the generalized velocity with of
rotational and translational components. As mentioned in Section
\ref{sec_metric}, the minimization solution is not invariant with
respect to changes in measure units if there is inhomogeneity of
units in the spaces of the force and the velocity
\cite{Lipkin-Duffy-1988,Manes-1992}. To go around the quandary of
inhomogeneous units, one can introduce a p.d. weight matrix by which
the coordinates of the force vector is changed to
\[ f_W = W^{-1/2} f \]
Note that the corresponding change of coordinates for the velocity
is $\dot{q}_W= W^{1/2} \dot{q}$ in order to preserve the
force-velocity product (by virtue of (\ref{eq_deltaqf})). Therefore,
the metric tensors for the force and velocity vectors are $W^{-1}$
and $W$, respectively. The inertia matrix and the Jacobian with
respect to the new coordinates are $M_W= W^{-1/2} M W^{-1/2}$ and
$A_W = A W^{-1/2}$. Since $\dot{q}_W$ and $f_W$ and the
corresponding projection matrix $P_W=I-A^+_WA_W$, where $A_W^+=[A
W^{-1/2}]^+$, is always dimensionless and, hence, invariant under
the measure units chosen. The new force and velocity vectors have
homogeneous units if the weight matrix is properly defined.
Therefore, replacing the new parameters, which are now dimensionally
consistent, in the optimal control (\ref{eq_fs}) minimizes $\| f_W
\|$ or equivalently minimizes the weighted Euclidean norm of the
generalized input force, i.e.,
\begin{equation} \label{eq_fWf}
\| f \|_W = (f^{T} W^{-1} f )^{1/2}.
\end{equation}
A quiet direct structure for $W$ is the diagonal one, i.e.,
\begin{equation} \label{eq_W}
W = \left[ \begin{array}{cc} \kappa^{-2} I & 0 \\ 0 &  I
\end{array} \right],
\end{equation}
where $\kappa$ is a length, by which we divide the
translational-velocity (or multiply the force). Using this length is
tantamount to the weighted norm as
\[ \| f \|_W = (\kappa^2 \| f_f \|^2 + \|
f_t \|^2  )^{1/2}, \] where the added terms are homogeneous, and
$f_f$ and $f_t$ are the force and the torque components of $f$. It
is worth pointing out that a {\em characteristic length}  that
arises naturally in the analysis and leads to invariant results was
proposed in \cite{Angeles-1997,Angeles-2003}.

Alteratively, the weight matrix can be selected according to some
engineering specifications. For instance, assume that the maximum
force and torque generated by the actuators are limited to
$f_{fmax}$ and $f_{tmax}$. Then, choosing $W^{1/2}=\mbox{diag}\{
f_{fmax},f_{tmax} \} $ leads to the minimization of this cost
function
\[ ( \| \frac {f_f}{f_{fmax}} \|^2 + \| \frac{f_t}{f_{tmax}} \|^2 )^{1/2}, \]
which, in a sense, takes the saturation of the actuators into
account.

\subsection{Control of Constraint Force}
\label{sec_ForceCntr}

The motion controller proposed in \ref{sec_MotionCntr} works well
for mechanical systems with bilateral constraint. On the other hand,
since the proposed  controller doesn't guarantee that the sign of
the constraint force will not change, a unilateral constraint
condition may not be physically maintained under the control law. In
this case, controlling the constraint force is a necessity.

Suppose that ${\cal F}_d$ represents desired constraint force which
can be derived from the desired Lagrange multipliers $\lambda_d$
using
\[ {\cal F}_d = A^T \lambda_d. \]
Then, considering $f_{\perp}^c$ as a control input, we propose the
following control law
\begin{equation} \label{eq_ForceContr}
f_{\perp}^c= h_{\perp}  + \mu(f_{\parallel} - h_{\parallel} +
C\dot{q}) + u_{\cal F},
\end{equation}
where $u_{\cal F}$ is the auxiliary control input which is
traditionally chosen as
\begin{equation} \label{eq_uf}
u_{\cal F} = {\cal F}_d  + G_F e_f + G_I \int_0^t e_f \mbox{d}
\tau,
\end{equation}
in which $e_f={\cal F}_d - {\cal F}$ is the force error, and $G_F>0$
and $G_I>0$ are the PI feedback gains. It should be pointed out that
the integral term is not necessary, but it improves the steady-state
error. From (\ref{eq_lamb}), (\ref{eq_ForceContr}), and
(\ref{eq_uf}), one can obtain the error dynamics as
\[ (G_F+1 ) \dot{e}_f + G_I e_f =0, \]
which will be stable provided that the gains are positive definite,
i.e., $ e_f \rightarrow 0 \;$ as $ \; t \rightarrow \infty$. Define
$e_{\lambda}=\lambda_d - \lambda$, and $e_f= A^T e_{\lambda}$. Then,
\begin{equation} \label{eq_elamb}
\underline{\sigma}(A) \| e_{\lambda} \| \leq  \| e_f \|,
\end{equation}
where $\underline{\sigma}(A)$ is the minimum singular value of the
Jacobian, i.e.,
\[ \underline{\sigma}(A) =  \;\;\left( \min_{eignvalue } A^T A \right)^{1/2}.
\]
Therefore, one can conclude  tracking of the Lagrange multipliers,
i.e. $ e_{\lambda} \rightarrow 0 \;$ as $ \; t \rightarrow \infty$,
if the Jacobian is full rank or $\underline{\sigma}(A) \neq 0$.

\begin{remark}
The constraint force ${\cal F}$ is always controllable, while the
Lagrange multipliers are controllable only if the Jacobian matrix is
of full rank.
\end{remark}

It is worth mentioning that, unlike the traditional motion/force
control schemes, which lead to coupled dynamics of force error and
position error, our formulation yields two independent error
equations. This is an advantage, because the motion control can be
achieved regardless of the force control and vice versa. To this
end, a hybrid motion/force control law can be readily obtained by
combining (\ref{eq_fs}) and (\ref{eq_ForceContr}),
\begin{eqnarray} \label{eq_hybrid}
f^c &=& f_{\parallel}^c + f_{\perp}^c \\ \nonumber &=& h + \mu C
\dot{q} + (I + \mu) P M u_p + u_{\cal F}.
\end{eqnarray}

\subsection{Control of Constrained Mechanical Systems with Passive
Joints} \label{sec_Passive}

Some constrained mechanical systems, e.g., parallel manipulators,
have joints without any actuators. The joints with and without
actuators are called {\em active joints} and {\em passive joints},
respectively. In this section, we use the notion of the linear
projection operator to generalize the inverse-dynamics control
scheme for constrained mechanical systems with passive joints.
Assuming there are $p$ active joints (and $n-p$ passive joints), the
generalized input force has to have this form
\[ f^c = \left[ \begin{array}{c} f^c_1 \\ \vdots \\ f^c_p \\ 0 \\ \vdots \\ 0
\end{array} \right] \!\!\!\!\!\!\!\!\!\!  \begin{array}{l} {\left. \begin{array}{c} \\ \\ \\ \end{array}
\right\} \mbox{active joints}} \\  {\left. \begin{array}{c} \\ \\ \\
\end{array} \right\}\mbox{passive joints} } \end{array}. \]
This implies that any admissible generalized force should satisfy
\begin{equation} \label{finR}
f^c \in {\cal R}(B)={\cal B}, \;\;\;\; \mbox{and} \;\;\;\; B =
\left[
\begin{array}{cc} I_p &  0 \\ 0 &  0
\end{array} \right],
\end{equation}
where $I_p$ is a $p\times p$ identity matrix. Note that $B$ is a
projection onto the {\em actuator space} ${\cal B}$, i.e., $B^2 = B$
and $B f^c =  f^c$.

Now, we need to modify the motion control law (\ref{eq_fs}) so that
the condition in (\ref{finR}) is fulfilled. If ${\cal N}(A)
\subseteq {\cal R}(B)$, then (\ref{finR}) is automatically satisfied
by choosing $f^c= f_{\parallel}^c$. Otherwise, we need to add a
${\cal N}^{\perp}$ component, say $f_{\perp}^c$, to
$f_{\parallel}^c$ so that $f^c= f^c_{\parallel}+ f^c_{\perp} \in
{\cal R}(B)$. Since $f_{\perp}^c$ does not affect the system motion
at all, the motion tracking performance is preserved by that
enhancement---albeit control of constraint force may no longer be
achievable. Let us assume
\begin{equation} \label{eq_fc}
f_{\perp}^c = (I - P) \eta,
\end{equation}
where $\eta \in \mathbb{R}^n$. Then, we seek $\eta$ such that
\begin{eqnarray} \nonumber
f^c \in {\cal R}(B) & \Leftrightarrow & (I - B) f^c =0
\\ \nonumber & \Leftrightarrow & (I - B)(f_{\parallel}^c
+(I- P) \eta )=0 \\ \label{eq_Qeta} & \Leftrightarrow & Q \eta = -(I
- B) f_{\parallel}^c
\end{eqnarray}
where $Q= I - B - P + B P$. Consider $\eta$ as the unknown variable
in (\ref{eq_Qeta}). A solution exits if the RHS of (\ref{eq_Qeta})
belongs to the range of $ Q$, i.e.,
\begin{equation} \label{eq_rangeQ}
{\cal R}((I - B) P) \subseteq {\cal R}((I-B)(I-P)).
\end{equation}
Then, the particular solution can be found via pseudo-inversion,
i.e.,
\begin{equation} \label{eq_eta}
\eta = - Q^+ (I - B) f_{\parallel}^c.
\end{equation}
The above equation yields the minimum-norm solution, i.e., minimum
$\| \eta \|$, which eventually minimizes the actuation force.
Equations (\ref{eq_fc}) and (\ref{eq_eta}) give
\begin{equation} f_{\perp}^c= H f_{\parallel}^c
\end{equation} where
\begin{equation} \label{eq_H}
H= -(I - P) Q^+ (I - B).
\end{equation}
Finally, we arrive at the following control law for constrained
mechanical systems with passive joints
\begin{equation} \label{eq_(I+H)f}
f^c = (I + H) f_{\parallel}^c,
\end{equation}
with $f_{\parallel}^c$ derived from (\ref{eq_fs}).

\subsubsection{Minimum-norm Torque}  A simple argument shows that the torque-control law
(\ref{eq_(I+H)f}), assuming the existence of a solution,  yields a
minimum-norm torque. Knowing that $\|I-P \|=1$, we have
\begin{eqnarray} \nonumber
\| f^c \|^2 &=& \| f_{\parallel}^c \|^2 + \|(I-P) \eta \|^2 \\
\label{eq_2norms} & \leq & \| f_{\parallel}^c \|^2 + \| \eta \|^2,
\end{eqnarray}
where both norms in the RHS of (\ref{eq_2norms}) are minimum.

\subsubsection{Controllability}
Because the existence of a solution  is tantamount to the
controllability condition of constrained mechanical systems under
the proposed control law, it is important to find out when a
solution to (\ref{eq_Qeta}) exists,  It can be inferred from
(\ref{eq_rangeQ}) that
\begin{equation}  \label{eq_ES}
\mbox{controllability cond.} \Leftrightarrow {\cal N} \cap {\cal
B}^{\perp} \subseteq {\cal N}^{\perp} \cap {\cal B}^{\perp}.
\end{equation}
In general, the proposed control method for systems with passive
joints works only if there exist sufficient number of active joints.
Since occurrence of the singularities gives rise to the number of
dof, the system under control law may no longer be controllable if
there are not enough active joints. For instance, mechanical systems
without any constraints at all cannot be controlled unless all
joints are actuated. This is because no-constraint means that
${\cal N}^{\perp} =\emptyset$ or ${\cal N} \in \mathbb{R}^n$; hence,
according to (\ref{eq_ES}), a controllable system requires that
$\mathbb{R}^n \cap {\cal B}^{\perp} \subseteq \emptyset \Rightarrow
{\cal B}^{\perp} = \emptyset$, which means there can be no passive
joints.

Also, it is worth pointing out that choosing ${\cal N} \subseteq
{\cal B}$ trivially results in a controllable system.

\section{A Slider-Crank Case Study}
\label{sec_Crank}
\begin{figure}
\centering{\includegraphics[width=7cm]{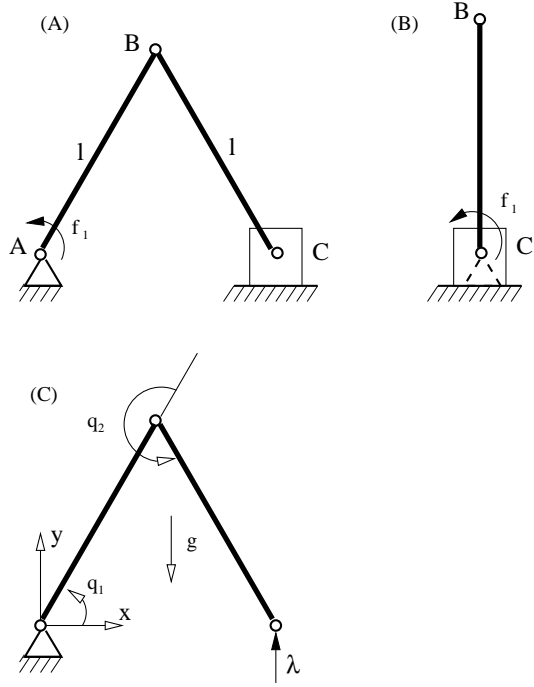}}
\caption{A slider-crank mechanism.} \label{fig_crank}
\end{figure}

In this section, we describe the results obtained from applying the
proposed inverse and direct dynamics formulations for simulation and
control of a slider-crank mechanism, Fig.\ref{fig_crank}A. Assume
that the dimension of the crank and the connecting rod are the same.
Then, singular configurations occur at
\begin{equation}\label{eq_singular}
q_1=\frac{n \pi}{2} \;\;\;\;\ n=\pm1, \pm2,\cdots,
\end{equation} see Fig.\ref{fig_crank}B.

\subsection{Equations of Direct Dynamics}

As shown in Fig.\ref{fig_crank}C, the closed-loop is cut in the
right hand support, i.e., at joint C. Let's vector $q=[q_1 \;
q_2]^T$ denote the joint angles. Then, the vertical position of the
point C is
\begin{equation} \label{eq_yC}
y_C(q) = l(s_1+s_{12}),
\end{equation}
where $l$ is the length of linkage. In the sequel,
$c_i$,$c_{12}$,$s_i$,$s_{12}$ are shorthanded for $\cos q_i$,
$\cos (q_1+q_2)$, $\sin q_i$, and $\sin (q_1+q_2)$ according to
the notation in \cite{Craig-1989}. The vertical translation motion
of point C is prohibited by imposing the following scleronomic
constraint equation
\begin{equation} \label{eq_yC=0}
\Phi(q) = y_C(q) =0
\end{equation}

Now, derivation of the direct dynamics may proceed as the following steps: \\

{\em Step 1:} Obtain the dynamic parameters of the open chain system
Fig.\ref{fig_crank}C (this is similar to the two-link manipulator
case study in \cite{Craig-1989}) as
\[ M =ml^2  \left[ \begin{array}{cc} 3+ 2 c_2 & 1 + c_2 \\
1 + c_2 & 1
\end{array} \right] \]
\begin{equation} \label{eq_Mh}
h= \left[ \begin{array}{c} -ml^2 s_2 (\dot{q}^2_2 + 2\dot{q}_1
\dot{q}_2)  + mlg (c_{12}+ 2c_1) \\
ml^2 s_2 \dot{q}^2_1 + mlg c_{12} \end{array} \right]
\end{equation}
where $m$ represent the mass of the link.

{\em Step 2:} Compute the projection matrix corresponding to the
constraint equation (\ref{eq_yC=0}). The Jacobian of the
constraint is
\begin{equation} \label{eq_A}
A= l\left[ \begin{array}{cc} c_1+c_{12} & c_{12} \end{array}
\right].
\end{equation}
The projection matrix can be computed numerically (e.g. using the
SVD of DSP Blockset in Matlab/Simulink \cite{Matlab_DSP}).
Nevertheless, we obtain $P$ in a closed form for this particular
illustration to have some insight into how the SVD handles
singularities. Since $q_2=2\pi- 2q_1$, the Jacobian matrix
(\ref{eq_A}) can be simplified as  $A= lc_1 [ 2 \;\; 1]$ whose
singular value is trivially $\sigma(A)=\sqrt{5}l |c_1|$. The SVD
algorithm treats all singular values less that $\epsilon$ as zeros,
i.e., $A^+=0$ if $|c_1| < \epsilon / l \sqrt{5}$. Hence
\begin{equation} \label{eq_PCrank}
P= \left\{ \begin{array}{cl} {\left[
\begin{array}{cc} 1/5 & -2/5 \\ -2/5 & 4/5 \end{array}  \right]} &
{\;\;\;\; \mbox{if} \;\;\;\; |c_1| < \frac{\epsilon}{l\sqrt{5}}}
\\ I & {\;\;\;\; \mbox{otherwise}}
\end{array} \right.
\end{equation}
This indicates the constraint is virtually removed if the system is
sufficiently closed to the singular configurations
(\ref{eq_singular}).

{\em Step 3:} Now, assuming $\gamma= ml^2$, one can compute the
constraint inertia matrix from (\ref{eq_M''c}) as
\begin{equation} \label{eq_McCrank}
M''_c = ml^2 \left[ \begin{array}{cc} 1 & (1+c_2)/5 \\
0 & (3-2c_2)/5
\end{array} \right] \;\;\;\; \mbox{if} \;\;\;\; |c_1| <
\frac{\epsilon}{l\sqrt{5}},
\end{equation}
and $M'_c = M$ at the singular positions---note that
$\det(M'_c)=m^2l^4(3-2c_2)/5 \neq0$. Also, note that $C \equiv 0$
for this example. Finally, plugging $P$ and $M'_c$ from
(\ref{eq_PCrank}) and (\ref{eq_McCrank}) into (\ref{eq_calMddq})
yields the direct dynamics of the slider-crank mechanism.

\section*{Conclusion}
\label{sec_conclusion}

A unified formulation  applicable to both the direct dynamics
(simulation) and inverse-dynamics (control) of constrained
mechanical systems has been presented. The approach is based on
projecting the Lagrangian dynamic equations into the tangent space
with respect to the constraint manifold. This automatically
eliminates the constraint forces from the equation, albeit the
constraint forces can be then retrieved separately from dynamics
projection into the normal space.

The novelty of the formulation lies in the definition of the
projector operators which, unlike in the other formulations, are
square matrices of order equal to the number of dependent
coordinates. Therefore, the structure of the dynamics formulation
doesn't change if the system changes its degree-of-freedom or its
topology. Moreover, since the process of computing projection
operator is not conditioned upon the maximal rank of the Jacobian,
the direct and the inverse-dynamics formulation are valid also for
mechanical systems with redundant constraints and/or singular
configurations, which is unattainable with many other classical
approaches.

A motion control system has been developed based on the {\em
projected inverse-dynamics control} which, minimizes actuation
force, and also works for systems having unactuated joints (passive
joints). To this end, a hybrid motion/force controller was
developed.

In summary, particular features of the proposed formulation
associated with simulation and control of constrained mechanical
systems are listed below:

\begin{itemize}
\item{A simulation may proceed even with presence of redundant
constraint equations and/or singular configurations. With the same
token, the projected inverse-dynamics motion controller can cope
with changes in the system's constraint topology or number of
degrees-of-freedom.}

\item{The generalized formulation requires no knowledge of the
constraint topology, i.e., description of how subassemblies are
connected, and it works for rigid-body or flexible systems alike.}

\item{The inverse-dynamics control scheme leads to minimum
weighted Euclidean norm of control force input.}

\item{The inverse-dynamics control scheme can be applied for
constrained systems which has some joints with no actuator.}

\item{If the inertia matrix posses a certain property as stated in
Proposition \ref{th_Decoupling}, the system exhibits decoupling
which leads to further simplification of the force control. }

\item{Redundant and flexible manipulators can be dealt with.}

\end{itemize}

\appendix

\section{}
\label{apx_RangeNull}
 Note that $A^T: \mathbb{R}^{m} \rightarrow
\mathbb{R}^{n}$. Then $\forall x \in {\cal R}(A^T)^{\perp}$ and
$\forall y \in \mathbb{R}^{m}$, we have
\begin{eqnarray} \nonumber
& & <x , A^T y > = 0      \\ \label{eq_ytAx} &\Leftrightarrow& x^T (
A^T y) = y^T A x =0 \;\;\;\;\; \forall y \in \mathbb{R}^{m}
\\ \label{eq_Ax} &\Leftrightarrow& A x =0
\\ \nonumber &\Leftrightarrow& x \in {\cal N}(A)
\end{eqnarray}
where the inference (\ref{eq_Ax}) is concluded because
(\ref{eq_ytAx}) implies that vector $A x$ must be orthogonal to
every vector in $\mathbb{R}^{m}$, and this is possible only if
vector $A x$ is identically zero. Thus (\ref{eq_rangA}) is proved.
The proof of (\ref{eq_rangAt}) can be shown by similar argument.

\section{}
\label{apx_Lagrang}
Assuming that the Lagrange multipliers is known, the acceleration
can be carried out from equation (\ref{eq_Mddq})
\begin{equation} \label{eq_ddq}
\ddot{q} = M^{-1}(f- h - A^T \lambda)
\end{equation}
Substituting the acceleration from (\ref{eq_ddq}) into the second
differentiation of the  constraint equation
\[ A \ddot{q} + \dot{A} \dot{q} =0 \]
gives the Lagrange multipliers as
\begin{equation} \label{eq_lambda}
\lambda= (A M^{-1} A^T)^{-1} \left[ A M^{-1}(f - h) + \dot{A}
\dot{q} + \dot{c} \right].
\end{equation}
This methods works only if the Cartesian inertia matrix ${\cal M}=(A
M^{-1} A^T)$ is not singular. Finally, substituting
(\ref{eq_lambda}) to (\ref{eq_ddq}) yields the acceleration.

\section{}
\label{apx_Invariant}

\begin{lemma} \label{inverse_invariant}
  The subspace ${\cal W}$ is invariant under an invertible
  transformation ${\cal A}$ if and only if the subspace is invariant
  under ${\cal A}^{-1}$.
\end{lemma}
{\sc Proof:}  By definition, ${\cal W}$ is an invariant subspace under
${\cal A}$ iff ${\cal A}{\cal W} \subseteq {\cal W}$. Moreover, the
invertible mapping ${\cal A}$ cannot reduce the dimension of any
subspace, that is $\mbox{dim}({\cal A}{\cal W}) = \mbox{dim}({\cal
  W})$ , hence ${\cal A}{\cal W} = {\cal W}$. It follows
\[ {\cal A}{\cal W}= {\cal W} \; \Leftrightarrow \; {\cal A}^{-1}{\cal W}= {\cal W}, \]
which completes the proof $\Box$.

\section{}
\label{apx_decoupling}
It is apparent from Fig.\ref{fig_constrained} that the decoupling
is achieved iff
\begin{equation} \label{eq_IPaP}
\mu P = (I - P) \alpha P = 0,
\end{equation}
which implies the null space must be invariant under $\alpha$. Since
$\alpha$ is an invertible mapping, it can be inferred from Lemma
\ref{inverse_invariant} (see Appendix \ref{apx_Invariant}) that the
null space must be invariant under $\alpha^{-1}$ too. Thais means
that (\ref{eq_IPaP}) is equivalent  to $(I - P) \alpha^{-1} P =0$.
Now, replacing $M_c$ from (\ref{eq_Mc}) into the latter equation and
after factorization, one can infer the followings
\begin{eqnarray} \nonumber
\mbox{decoupling} &\Leftrightarrow& (I - P) [(M+ \tilde{M}) M^{-1} ] P =0  \\
\nonumber
 & \Leftrightarrow& [(I - P)M P][M^{-1} P] =0 \\ \nonumber
 & \Leftarrow& (I - P) M P =0
\end{eqnarray}
which completes the proof.

\section{}
\label{apx_rank}

Since $\theta^T=[\theta_1(q) , \cdots, \theta_k(q)]$ is a set of
independent functions, the corresponding Jacobian matrix is full
rank, i.e., $\mbox{rank}( \frac{\partial \theta}{ \partial q}) =k$.
Moreover, by using the chain-rule, we have $\Lambda \frac{\partial
\theta}{\partial q} = I_{k}$, where $I_{k}$ is a $k \times k$
identity matrix. Now, by virtue of the property of the rank operator
\footnote{$\mbox{rank}(A B) \leq \min [ \mbox{rank}(A) ,
\mbox{rank}(B)]$}, one can say
\[ \min[ \mbox{rank}(\Lambda) , k ] \leq k, \]
or
\[ \mbox{rank}(\Lambda) = k. \]

\bibliographystyle{IEEEtran}

\end{document}